\documentclass[twocolumn,showpacs,preprintnumbers,amsmath,amssymb,aps,prb]{revtex4}
\usepackage{graphicx}
\begin{document}

\title{
Quantized Transport for a Skyrmion Moving on 
a Two-Dimensional Periodic Substrate 
} 
\author{
C. Reichhardt, D. Ray, and C. J. Olson Reichhardt 
} 
\affiliation{
Theoretical Division and Center for Nonlinear Studies,
Los Alamos National Laboratory, Los Alamos, New Mexico 87545 USA\\ 
} 

\date{\today}
\begin{abstract}
We examine the dynamics of a skyrmion moving over a two-dimensional periodic 
substrate utilizing 
simulations of a particle-based skyrmion 
model. We specifically examine the role of 
the non-dissipative Magnus term on the driven motion 
and the resulting skyrmion velocity-force curves.    
In the overdamped limit, 
there is a  depinning transition 
into a sliding state in which the skyrmion moves 
in the same direction as the external drive. 
When there is a finite Magnus component in the equation of motion,  
a skyrmion in the absence of a substrate
moves at an angle 
with respect to the direction of the external driving force. 
When a periodic substrate is added,
the direction of motion or Hall angle of the skyrmion is dependent 
on the amplitude of the external drive,   
only approaching the substrate-free limit for higher drives. 
Due to the underlying  
symmetry of the substrate the 
direction of skyrmion motion does not 
change continuously as a function of drive, but rather forms
a series of discrete steps corresponding to 
integer or rational ratios of the
velocity components perpendicular
($\langle V_{\perp}\rangle$) and parallel 
($\langle V_{||}\rangle$) to 
the external drive direction: $\langle V_{\perp}\rangle/\langle V_{||}\rangle=n/m$,
where $n$ and $m$ are integers. 
The skyrmion passes through
a series of directional locking phases 
in which the motion is locked to 
certain symmetry directions of the substrate for 
fixed intervals of the drive amplitude. Within a 
given directionally locked phase, 
the Hall angle remains constant and the 
skyrmion moves in an orderly fashion through the sample.    
Signatures of the transitions into and out of these locked phases 
take the form of
pronounced cusps in the skyrmion velocity versus force curves, as well as 
regions of negative differential mobility 
in which the net skyrmion velocity 
decreases with increasing external driving force.
The number of steps in the transport curve increases when 
the relative strength of the Magnus term 
is increased. 
We also observe
an overshoot phenomena 
in the directional locking, 
where the skyrmion motion can lock to a Hall angle 
greater than the clean limit value and then 
jump back to the lower value at higher drives.
The skyrmion-substrate interactions can also 
produce a skyrmion acceleration effect in which,
due to the non-dissipative dynamics, 
the skyrmion 
velocity exceeds the value expected to be produced by the
external drive.
We find that these effects are robust for different types of periodic
substrates.  
Using a simple model for a skyrmion interacting with a single pinning site, 
we can capture   
the behavior of the change in the Hall angle with increasing external drive.  
When the skyrmion moves through the pinning site, its trajectory
exhibits a side step phenomenon 
since the Magnus term induces
a curvature in the skyrmion orbit. 
As the drive increases, this curvature is reduced 
and the side step effect is also reduced.   
Increasing the strength of the Magnus term reduces the 
range of impact parameters over which the skyrmion can be
captured by a pinning site,
which is one of the reasons that strong Magnus force effects reduce 
the pinning in skyrmion systems.   
\end{abstract}
\maketitle

\section{Introduction}
Skyrmions 
were predicted to occur in certain magnetic systems \cite{1} 
and were subsequently experimentally identified in the 
chiral magnet MnSi \cite{2}. 
Since this initial discovery 
there has been tremendous growth in the field as an increasing number of    
materials have been found that 
can support a skyrmion phase \cite{3,4,5,6,7,8,9,10}. 
There are also numerous 
proposals on how 
to stabilize skyrmion states by utilizing 
different materials properties or bilayers 
\cite{11,12,13,14}.       
Direct imaging of skyrmions with
Lorentz microscopy \cite{3,4,5,7,10}  
and other techniques \cite{8,15,16} show that
the skyrmions form a triangular lattice 
and have particle-like properties similar 
to vortices in type-II superconductors \cite{38}.
As an external magnetic field is increased,
skyrmions emerge from a spiral state and their density
initially increases and then decreases with field until the
sample
enters a uniform ferromagnetic state \cite{2,3,9}. 
In bulk samples, skyrmions form
three-dimensional (3D) line objects and occur in a limited
range of fields and temperatures \cite{2,16},  
while for thin samples the skyrmions 
exhibit two-dimensional (2D) properties and are
stable over a much larger range of fields and temperatures extending
close to room temperature
\cite{4,5,6,7}. 
Skyrmions can be set into motion  
through the application of an external current 
\cite{17,7,18,19,10}, and it has been shown that 
there is a critical current 
above which the skyrmions depin into a sliding state 
\cite{18,19,21,23,24,25,26,27}. 
Skyrmion motion can be directly observed with Lorentz microscopy
\cite{7,10} 
or deduced from changes in the 
transport properties, permitting the construction of
effective skyrmion velocity versus applied force curves that show that
the skyrmion
velocity increases with increasing current \cite{19}.
Other methods to move skyrmions include the use 
of temperature gradients \cite{28,29,30,31,32}, 
electric fields \cite{33,34}, and coupling to a magnetic tip \cite{15}.         

From an applications standpoint,
skyrmions are attracting attention 
due to their potential use 
in racetrack memory devices
where they would play 
a role similar to that of  
magnetic domain walls \cite{35,36,37}. 
Skyrmions have several advantages over domain walls  
due to their size and the fact that
the current needed to depin a skyrmion 
can be orders of magnitude smaller than that needed to move domain walls
\cite{19,35}. 
It has been experimentally demonstrated that individual skyrmions 
can be created or 
annihilated with a magnetic tip, indicating that it is feasible
to read and write skyrmions \cite{16}.
Developing applications of skyrmions will require an
understanding of how skyrmions 
interact with and move along 
tailored landscapes, so examining skyrmion dynamics 
on periodic substrates is an 
important step in this direction.  

Skyrmions
have many similarities to vortices in type-II superconductors,
such as
the effective skyrmion-skyrmion
interaction, which is repulsive and favors triangular ordering, 
and the fact that both skyrmions and vortices 
can be driven by an external current.
In the presence of quenched disorder, skyrmions 
exhibit pinning-depinning transitions \cite{37},
as observed in experiments \cite{19} and simulations \cite{21,23,24,25,26}, 
and these are
similar to the depinning transitions observed 
for vortices in type-II superconductors \cite{38,39,40,41}. 
There are important differences between the two systems, 
particularly the dominant role that non-dissipative effects can
take in skyrmion motion.
For superconducting vortices, non-dissipative effects such as a Magnus force
are typically small, permitting
the system to
be effectively described as obeying overdamped dynamics \cite{38,40}. 
In a skyrmion system, by contrast,
the Magnus term can
strongly affect 
how the skyrmions interact with pinning sites and 
how they move under an external drive \cite{18,21,23}. 
Numerical simulations using continuum and particle-based models for skyrmions
interacting with pinning have shown that the Magnus term reduces the
effective pinning in the system by creating
a velocity component that is perpendicular to the force 
induced by a 
pinning site \cite{18,21,23,24,26}. 
As a result,
skyrmions have a tendency to swing around the edge of a pinning site
and escape, while for overdamped
systems a particle moves toward the center of a pinning site and
is much more likely to be pinned.      

Relatively little is known about how particles with  a strong Magnus 
term move over a periodic substrate,  
and skyrmions are an ideal system to study such effects.
In certain limits, a skyrmion can be effectively modeled as a point-like 
particle utilizing an equation of motion \cite{23,24} 
derived from Thiele's equation \cite{42}.  
Comparison between 
continuum-based models and particle-based models of skyrmions moving in
the presence of pinning have shown good agreement \cite{23,24}. 
In this work we examine the dynamics of 
a skyrmion moving over a  
square periodic substrate using a particle-based description
given in Section II.  
We specifically examine the effect 
of changing the importance of the Magnus term relative to that 
of the damping term.
Despite the apparent simplicity of this 
system, we show in Section III that the Magnus term 
can induce a remarkably rich variety of dynamical behaviors 
that are absent in the overdamped limit.
We find that the Hall angle 
for the skyrmion motion is dependent on the external drive amplitude 
and approaches the substrate-free limit only
at higher drives.  
Since the skyrmion is moving over a periodic substrate,
as the Hall angle changes the 
motion becomes locked to specific symmetry directions of the 
substrate, producing a series
of steps in the transport curves 
corresponding to integer and rational fractional
ratios of the skyrmion velocity in the directions parallel and
perpendicular to the drive direction.
At the
transitions into the different directional locking phases, 
the skyrmion velocity exhibits 
a pronounced cusp accompanied by  
negative differential mobility in which the skyrmion velocity {\it decreases} 
with increasing external driving force.
We map the extent of the locking phase as a function
of the external drive and the ratio of the Magnus and dissipative terms, 
and find a rich structure of integer and fractional locking effects.
We also describe a speedup effect for skyrmions interacting with a
substrate
where the skyrmion
velocity can be higher than the external drive. 
This effect
is most prominent just above depinning and is 
caused by the Magnus term; it is absent in the overdamped case.  
In Section IV we consider a skyrmion scattering from a single pinning site 
for varied impact parameters to show     
how the Hall angle is reduced by the pinning due to   a
side-step phenomenon where the skyrmion trajectory is 
shifted by the Magnus term as the skyrmion moves
through the pinning site.
For increasing drive amplitude, the skyrmion trajectories become 
less curved and the size of the
side step is reduced, so that the Hall angle approaches 
the clean limit for higher drives.  
Our results for the speedup effect and side steps are 
in agreement with recent theoretical and computational studies 
by M{\" u}ller and Rosch, who considered a single skyrmion 
interacting with a defect site \cite{27}.
In that work, the pinning potential is of a different form than the pinning
sites we consider;
however,
the consistency of the two studies  indicates
that the Hall angle dependence on external drive and 
the speedup phenomenon 
are generic features of skyrmions interacting with pinning. 

There are other examples of particles 
moving over ordered substrates, 
such as vortices in type-II superconductors
with periodic pinning arrays \cite{43,44,45}
or colloids placed on optically created periodic  substrates \cite{47,48}. 
In these systems the dynamics is overdamped;
however, there can be 
directional locking 
effects in which the particles preferentially move along 
symmetry directions of the underlying substrate
as the direction of drive is rotated with respect to 
the substrate lattice \cite{51,52,53,54,55,56,57,58}. 
Such directional locking effects can be exploited to perform
particle separation in colloidal systems by setting up a system in which
one particle species locks to the substrate while the second species
does not, causing the two species to move at an angle with each
other
\cite{53,54}. 
The direction of motion of the locked particles undergoes a
series of steps
as a function of the effective angle of drive 
with respect to the substrate.  The steps are 
centered at integer and
rational ratios of the angle of drive 
and form a devil's staircase structure \cite{52,53,54,56,57,59,60}.
In the skyrmion system, we observe directional locking effects
when the direction of external drive is {\it fixed} 
with respect to the substrate and the drive amplitude is varied. 
The directional locking and transitions into the locking states 
in the skyrmion system exhibit a number of features that have not 
been observed
in overdamped systems, 
such as an overshoot effect in the locking direction 
and negative differential mobility
at the locking transition.  
The steps in the transport response are also
distinct from the Shapiro steps found in systems 
that can be effectively described as 
a particle moving over a periodic substrate under superimposed ac and
dc drives
\cite{61,62}. In the system we consider in this work,
there is no imposed ac drive.    

\section{Simulation and System}   

We consider a skyrmion moving over a 2D square periodic substrate  
and utilize a particle-based description of the skyrmion from a recently 
developed equation of motion for skyrmions 
\cite{23,24}. 
The dynamics of the
skyrmion is obtained by integrating the equation of motion:
\begin{equation}  
\alpha_d\frac{d {\bf R}_{i}}{dt} = 
{\bf F}^{s}_{i} + \alpha_m {\hat {\bf z}}\times {\bf v}_{i} +  {\bf F}^{D}_{i}  \ .
\end{equation} 
Here $\frac{d {\bf R}_{i}}{dt} = {\bf v}_{i}$ is
the skyrmion velocity,  and $\alpha_d$ is a damping term 
representing spin precession and dissipation of electrons 
localized in the skyrmion. 

The substrate force 
${\bf F}^{s}_{i}$ is applied by placing $N$ skyrmions in a square lattice
with lattice constant $a$ and fixing them in place.
These 
skyrmions interact repulsively with the mobile skyrmion.
Such a substrate could be created experimentally
by placing an array of magnetic dots on the sample to create
a background of effectively immobile skyrmions. 
Alternatively,
in recent continuum-based simulations it was shown that 
a defect site formed by removing a single spin creates 
a potential with a long range repulsion 
and a short range attraction \cite{27}.
Our system can then be viewed as containing a periodic array 
of such pinning sites in the limit where the moving skyrmion
does not experience the shorter range attraction 
of the pins. 
We also consider a model in which the substrate potential is
represented by a 2D sinusoidal array and find behavior very similar
to that of a skyrmion moving through a pinned skyrmion lattice,
indicating that our results are robust and capture the general
features of skyrmions moving over periodic substrates.
The force from the pinned skyrmions has the form
${\bf F}_i^s=\sum^N_{j = 1}K_{1}(R_{ij} /\xi){\hat {\bf R}}_{ij}$, 
where $R_{ij}$ is the distance between the driven skyrmion $i$ and 
an immobile
skyrmion $j$. 
Here $K_{1}$ is the modified Bessel function, 
which falls off exponentially for large $R_{ij}$, 
and $\xi$ is a screening length which 
we take to be $1.0$ in dimensionless units. 
The sample size is $L \times L$ with $L=36$, and the lattice
constant of the substrate is $a=3.26$.
In the second system we model the substrate with the 2D periodic form
${\bf F}_{s}^i = F_p[\cos^2(\pi x/a){\bf {\hat x}} +  \cos^2(\pi y/a){\hat {\bf y}}]$, where $a=3.26$ is the substrate lattice 
constant and $F_p$               
is the amplitude of the substrate force. 
In both systems we employ periodic boundary conditions.   

The Magnus term 
$ {\bf F}^{M} =\alpha_m {\hat {\bf z}}\times {\bf v}_{i}$  
produces 
a force that is perpendicular to the skyrmion velocity, 
where $\alpha_m$ is the magnitude of the Magnus term.  
The driving force is
${\bf F}_i^{D}=F_D{\bf \hat d}$ where ${\bf \hat d}$ is the direction of 
the applied drive.  Such a force
could arise due to the application of an external
current to the skyrmion \cite{23,24}.
In most of this work, we take ${\bf \hat d}={\bf \hat x}$.
We measure the skyrmion 
velocity parallel, $\langle V_{||}\rangle$, and perpendicular,
$\langle V_{\perp}\rangle$, to the drive.
In the absence of a substrate  
or in the overdamped limit $\alpha_m/\alpha_d = 0.0$,
the skyrmion moves in the direction of the drive, 
while for a finite $\alpha_m/\alpha_d$ the 
skyrmion moves at an angle $\Theta$ with respect to the drive,
where
$\Theta = \arctan(\langle V_{\perp}\rangle/\langle V_{||} \rangle) 
= \arctan(\alpha_m/\alpha_d)$. 
Increasing $\alpha_m/\alpha_d$ produces a larger angle  
$\Theta$ for the skyrmion motion with respect to the external drive. 
To quantify the direction of motion 
we measure $R = \langle V_{\perp}\rangle/\langle V_{||}\rangle$ so that
the Hall angle $\Theta$ is given by $\arctan(R)$.
We increase the external drive in small increments of 
$\delta F_D=0.001$ to $0.005$ and wait several thousand 
simulation time steps before
measuring the average velocity in order to ensure that the 
skyrmion velocity is in a steady state. We find
that for smaller increments $\delta F_D$ our results do not change.    
Throughout this work we impose the constraint $\alpha_d^2+\alpha_m^2=1$ in 
order to maintain a constant magnitude of the skyrmion velocity for
varied ratios of $\alpha_m/\alpha_d$.

\begin{figure}
\includegraphics[width=3.5in]{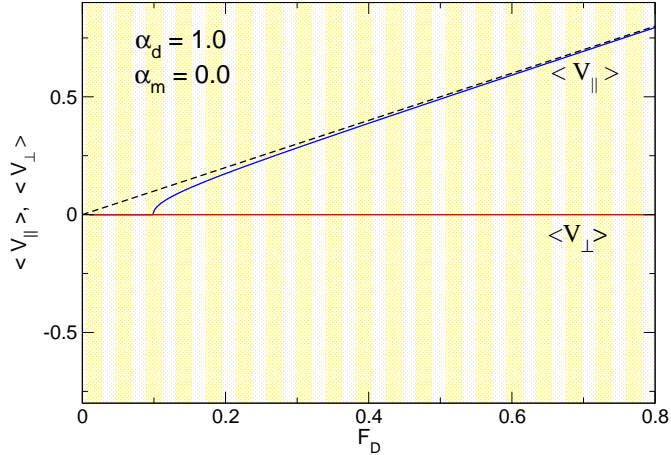}
\caption{ 
(a) The velocity for a skyrmion under a dc driving force $F_{D}$ 
in the overdamped
limit with $\alpha_m/\alpha_d = 0$ moving over a periodic substrate. 
The drive is applied
in the $x$-direction; the system geometry is illustrated in Fig.~\ref{fig:2}. 
$\langle V_{||}\rangle$ is the velocity component 
parallel to the applied drive and $\langle V_{\perp}\rangle$ 
is the velocity component perpendicular to the drive. 
Here there is a depinning transition into a sliding state where the 
skyrmion moves strictly in the direction
of the applied drive.  The Hall term 
$R = \langle V_{\perp}\rangle/\langle V_{||}\rangle = 0.0$ in
this case.     
Dashed line: $\langle V_{||}\rangle$ response in the
absence of a substrate.
}
\label{fig:1}
\end{figure}

\begin{figure}
\includegraphics[width=3.5in]{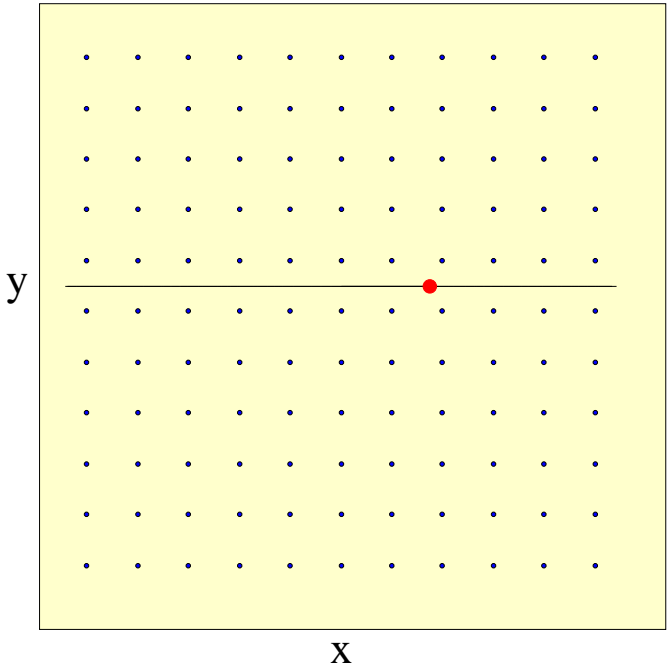}
\caption{The geometry for the system in Fig.~\ref{fig:1} 
with $\alpha_m/\alpha_d=0$
where the skyrmion (large red dot) is driven in the $x$-direction 
under an applied drive $F_{D}$. 
The black line is the skyrmion trajectory and the smaller blue dots
indicate the locations of the potential maxima in the periodic 
square substrate lattice.
}
\label{fig:2}
\end{figure}

\section{Velocity-Force Curves and Directional Locking}

In Fig.~\ref{fig:1} we plot $\langle V_{||}\rangle$ 
and $\langle V_{\perp}\rangle$ 
versus $F_{D}$ for a skyrmion moving
over a periodic substrate in the  
overdamped case where $\alpha_d = 1.0$ and $\alpha_m = 0.0$. 
Figure \ref{fig:2} shows the
system geometry, highlighting the motion of the mobile skyrmion through
the background of potential maxima.
In Fig.~\ref{fig:1} there is a depinning transition to a sliding
state for $F_D>0.1$ as indicated by $\langle V_{||}\rangle>0$.
Here the skyrmion moves strictly in the direction of the applied
drive so that $\langle V_{\perp}\rangle=0$ for all $F_D$ and
$R = \langle V_{\perp}\rangle/\langle V_{||}\rangle = 0$.   
The dashed line in Fig.~\ref{fig:1} is the expected 
value of $\langle V_{||}\rangle$ in the clean limit, and we find that
as $F_D$ increases,
$\langle V_{||}\rangle$ gradually approaches the clean limit value.    
Figure~\ref{fig:2} shows the  
skyrmion trajectory for fixed $F_{D} = 0.5$, where motion occurs 
in a straight line along the driving direction. 

\begin{figure}
\includegraphics[width=3.5in]{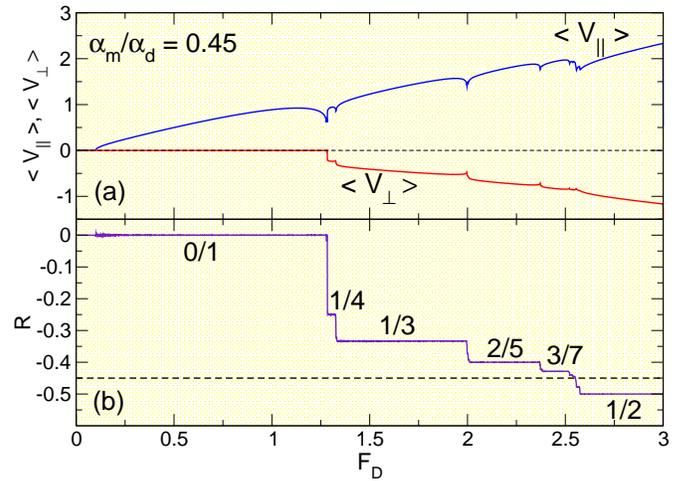}
\caption{(a) $\langle V_{\perp}\rangle$ and $\langle V_{||}\rangle$ 
vs $F_{D}$ for the same system as in Fig.~\ref{fig:1} but with
$\alpha_m/\alpha_d = 0.45$. (b) $R$ vs $F_{D}$ in the same sample.
The dashed line indicates that in the clean limit $R=-0.45$.
There are a series of dips in the velocity-force curves correlated with
jumps in $R$, indicating 
that the skyrmion undergoes a series of transitions between 
locking to different symmetry directions of the substrate.
The largest steps in $R$, corresponding to $n/m$ ratios of
$0/1$, $1/4$, $1/3$, $2/5$, $3/7$,
and $1/2$, are marked;
there are also
additional smaller higher order steps at other integer values of $n$ and $m$.
}
\label{fig:3}
\end{figure}

\begin{figure}
\includegraphics[width=3.5in]{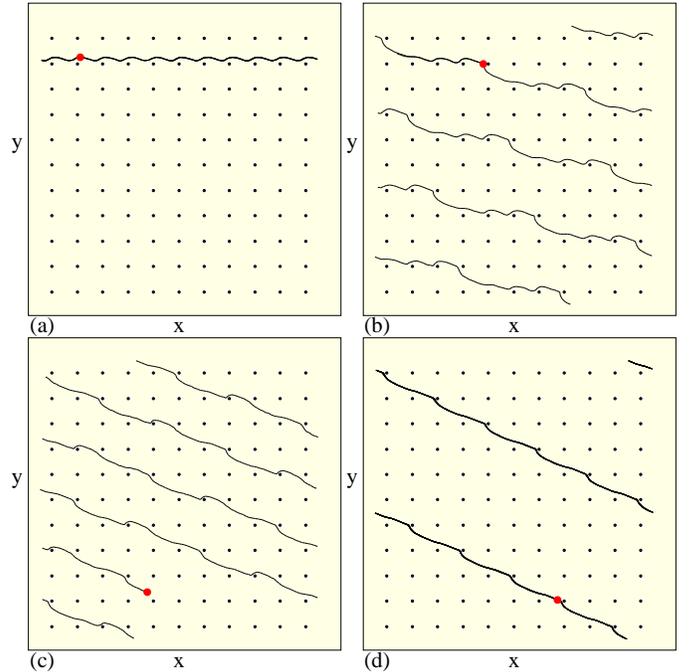}
\caption{The skyrmion trajectories and substrate maxima for the system in 
Fig.~\ref{fig:3} with $\alpha_m/\alpha_d = 0.45$.
(a) $|R|=0/1$ state at $F_{D} = 1.0$. (b) $|R|=1/4$ state at $F_{D} = 1.3$. (c) 
 $|R|=2/5$ state at $F_{D} = 2.2$. (d) $|R|=1/2$ state at $F_{D} = 2.75$.  
}
\label{fig:4}
\end{figure}

\subsection{Directional Locking}

In Fig.~\ref{fig:3} we illustrate the transport behavior in a sample with
$\alpha_m/\alpha_d = 0.45$, $\alpha_m = 0.41$, and $\alpha_d = 0.912085$. 
In the absence of a substrate the skyrmion would move at an angle 
of $|\Theta| = 24.227^{\circ}$ with respect to the drive direction 
and would have $R=\langle V_{\perp}\rangle/\langle V_{||}\rangle = -0.45$.  
Figure~\ref{fig:3}(a) shows $\langle V_{||}\rangle$ and 
$\langle V_{\perp}\rangle$ vs $F_{D}$, 
and in Fig.~\ref{fig:3}(b) we plot $R$ vs $F_{D}$.  Here 
there is a transition from a pinned to sliding state just above 
$F_{D} = 0.1$. 
The skyrmion moves strictly in the direction of the drive for 
$0.1 < F_{D} < 1.28$, which corresponds to the $R = 0$ regime 
shown in Fig.~\ref{fig:3}(b) for the same interval of $F_D$.
Figure~\ref{fig:4}(a) shows the skyrmion trajectory in the $R = 0$ region 
for $F_{D} = 1.0$ for the system in Fig.~\ref{fig:3}(a). 
The skyrmion no longer passes through the centers of the potential 
minima but its trajectory shifts
closer to 
the bottom row of potential maxima and 
develops an oscillation in the $y$ direction
that was absent in the $\alpha_m/\alpha_d = 0$ case shown in Fig.~\ref{fig:2}.
As $F_{D}$ is further increased the skyrmion shifts even
closer to the row of potential maxima
and its velocity parallel to the drive decreases,
as indicated by the dip in $\langle V_{||}\rangle$ near $F_{D} = 1.28$.
As $F_D$ increases further, both $\langle V_\perp\rangle$ and
$\langle V_{||}\rangle$ increase, indicating that the skyrmion is
now translating in the $y$ direction as well as in the $x$ direction.
Additional dips in both $\langle V_{||}\rangle$ and $\langle V_{\perp}\rangle$ 
occur at higher $F_D$, 
with occasional regions containing
multiple closely spaced smaller dips. 
Figure~\ref{fig:3} indicates
that these dips correlate with the jumps in  
$R$, and that between the dips, $R$ remains constant. 
This means that the direction of motion or Hall angle of the
skyrmion changes in a discrete fashion with increasing $F_{D}$. 
The steps in $R$ appear at values 
that are rational ratios of 
$\langle V_{\perp}\rangle/\langle V_{||}\rangle$ 
of the form $n/m$, where $n$ and $m$ are integers.
In Fig.~\ref{fig:3}(b) we highlight the steps at 
$|R| = 1/4$, 1/3, 2/5, 3/7, and $1/2$. There are also
numerous additional   
steps in $R$ for smaller intervals of $F_{D}$ for higher order
rational ratios of $R = n/m$. In general,
the larger the values of $n$ and $m$, the smaller the interval in 
$F_{D}$ over which the step appears.

During each step interval in $R$,
the skyrmion follows an ordered periodic orbit,
translating $n$ substrate plaquettes in the direction perpendicular to
the drive for every $m$ plaquettes it translates in the direction
parallel to the drive.
Figure~\ref{fig:4}(a) shows the orbit of a 
skyrmion in the  $|R|=0/1$ state, 
while Fig.~\ref{fig:4}(b) illustrates the orbit 
in the $|R| = 1/4$ state at $F_{D} = 1.3$. 
Here the skyrmion moves periodically through the system
at an angle 
$\Theta = \arctan{1/4} = 14.036^\circ$
with respect to the drive direction. 
In each period of the motion,
the skyrmion translates to the right
by four plaquettes in the drive or $x$ direction and
down by one plaquette in the perpendicular or $y$ direction,
giving $n/m = 1/4$.
At the transition out of the $|R| = 1/4$ state, the skyrmion
slows down, producing a cusp in the velocity curves in 
Fig.~\ref{fig:3}(b) 
and a dip in the net velocity near $F_{D} = 1.327$.  
A similar ordered orbit occurs
in the $|R|= 1/3$ state, where the skyrmion 
moves three plaquettes in the $x$-direction and one plaquette in 
the $y$ direction during every period.
There is another dip in the net skyrmion velocity
near $F_{D} = 2.0$ that occurs when
the system transitions to the $|R| = 2/5$ state.
Figure~\ref{fig:4}(c) illustrates the
orbit at this value of $R$ for $F_{D} = 2.2$. 
Here the periodic orbit is more extended and the skyrmion 
moves two plaquettes in the 
negative $y$-direction for every
five plaquettes in the $x$-direction.
Above the $|R| = 2/5$ state, the system 
enters the $|R| = 3/7$ state, followed by some 
higher order steps which occur over small intervals of $F_{D}$ as 
shown in  
Fig.~\ref{fig:3}(a) near $F_{D} = 2.55$. 
The system then reaches the $|R|= 1/2$ state illustrated in
Fig.~\ref{fig:4}(d) 
at $F_{D} = 2.75$, where 
the skyrmion moves two plaquettes in the $x$-direction 
for every one plaquette in the $y$-direction. 
At the other higher-order locking steps, similar ordered orbits appear.

The dips in $\langle V_{||}\rangle$ and $\langle V_{\perp}\rangle$ 
in Fig.~\ref{fig:3}
are associated with transitions in the skyrmion orbit 
from one directionally locked state to another, 
with a corresponding change in the Hall angle. 
The symmetry of the square lattice determines the specific
directions along which the skyrmion motion locks, so that
for other geometries such as a triangular substrate, the locking
directions will be different.
At the transitions between locking steps, the net skyrmion velocity  
$\langle V\rangle = ( \langle V_{||}\rangle^2 + \langle V_{\perp}\rangle^2)^{1/2}$ 
decreases with increasing external drive $F_{D}$. 
This phenomenon is known 
as negative differential mobility, and 
it has been observed in other systems where particles are driven over a
periodic substrate, such as 
superconducting vortices moving over periodic
pinning arrays where there are transitions between different dynamical 
phases \cite{45}. 
Negative differential mobility is also a common feature in 
semiconductor devices \cite{63} and can be useful for creating 
logic devices. 
The ability to 
control differential mobility
in skyrmion systems could open new approaches for applications.      

\begin{figure}
\includegraphics[width=3.5in]{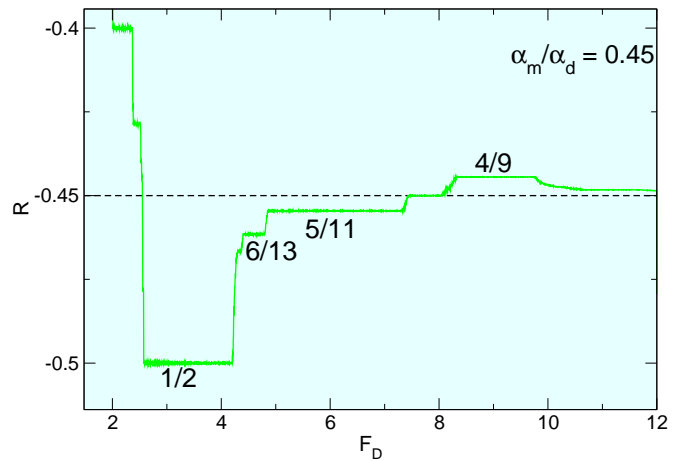}
\caption{A detailed view of the $R$ vs $F_{D}$ curve for the 
same system in Fig.~\ref{fig:3} at $\alpha_m/\alpha_d = 0.45$,
where we examine the locking for drives up to $F_{D} = 12.0$. 
The system is locked to the $|R| = 1/2$ value but then 
jumps to $|R| = 6/13$ and $5/11$ before reaching $|R| = 4/9$. 
As $F_{D}$ is further increased,
$R$ gradually approaches the clean limit value, indicated by the dashed line.  
}
\label{fig:5}
\end{figure}

\subsection{Overshoot Effect at $\alpha_m/\alpha_d=0.45$}

At $F_{D} = 3.0$
in Fig.~\ref{fig:3}(b), $|R|=0.5$, indicating
the skyrmion is
moving at a Hall angle of 
$\Theta = 25.565^\circ$, which is 
{\it higher} than the clean limit value 
of $|R| = 0.45$ or $\Theta = 24.23^\circ$. 
This phenomenon occurs when the clean value of $\Theta$ is oriented
close to but slightly below
a strong symmetry locking direction of the substrate.  The skyrmion locks
to the substrate and its Hall angle slightly exceeds that expected
for the clean limit.
As $F_{D}$ is increased, the effectiveness of the substrate 
is reduced and the direction of motion of the skyrmion
gradually approaches the clean limit value, 
as shown in Fig.~\ref{fig:5} where  
we plot a portion of the $R$ vs $F_{D}$ curve for the same 
system from Fig.~\ref{fig:3}(b) but for 
drives up to a higher value of $F_{D} = 12.0$. 
The dashed line in Fig.~\ref{fig:5} indicates the clean value limit 
of $|R| = 0.45$.  
For $F_{D} > 4.2$, there is a jump from the $|R| = 0.5 = 1/2$ state 
to a lower value of $|R| = 0.46143 = 6/13$, followed by another jump 
to the $|R| = 0.4545 = 5/11$ state. 
There is then a small region
where the system locks to the $|R| = 0.45=9/20$ state before jumping to 
$|R| = 0.444 = 4/9$.  
As $F_D$ increases further,
$R$ gradually
approaches the clean limit value, 
as indicated by the dashed line
in Fig.~\ref{fig:5}(b).
This shows that there can be an overshoot effect in the locking
behavior
for a certain range of drives,   
giving rise to non-monotonic behavior of $R$ as a function of  $F_{D}$. 
We have observed similar overshoot effects for other values of  
$\alpha_m/\alpha_d$.          

\begin{figure}
\includegraphics[width=3.5in]{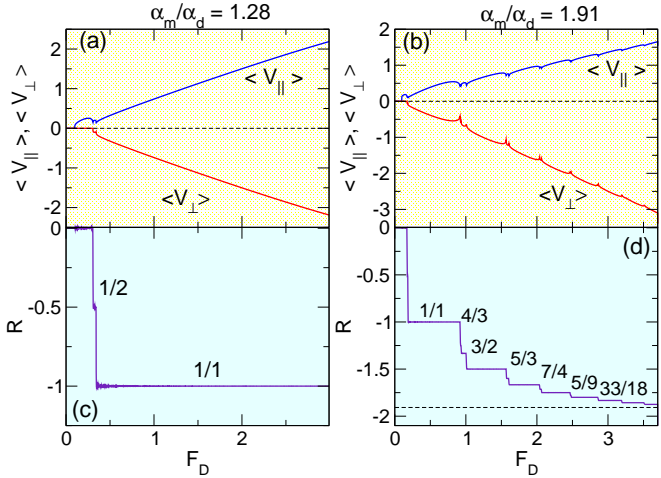}
\caption{ 
(a,b) 
$\langle V_{\perp}\rangle$ and $\langle V_{||}\rangle$ 
vs $F_{D}$ for (a) $\alpha_m/\alpha_d = 1.28$ and (b) $1.91$.
(c,d) The corresponding $R$ vs $F_{D}$ curves.  
Dashed line in (d): the clean limit value for $R$.
In (c) we highlight the $|R|=0/1$ and $|R|=1/1$ steps, 
and in (d) we highlight the 
$|R| = 1/1$, 4/3, 3/2, 5/3, 7/4, $5/9$ and $33/18$ steps.   
}
\label{fig:6}
\end{figure}

\begin{figure}
\includegraphics[width=3.5in]{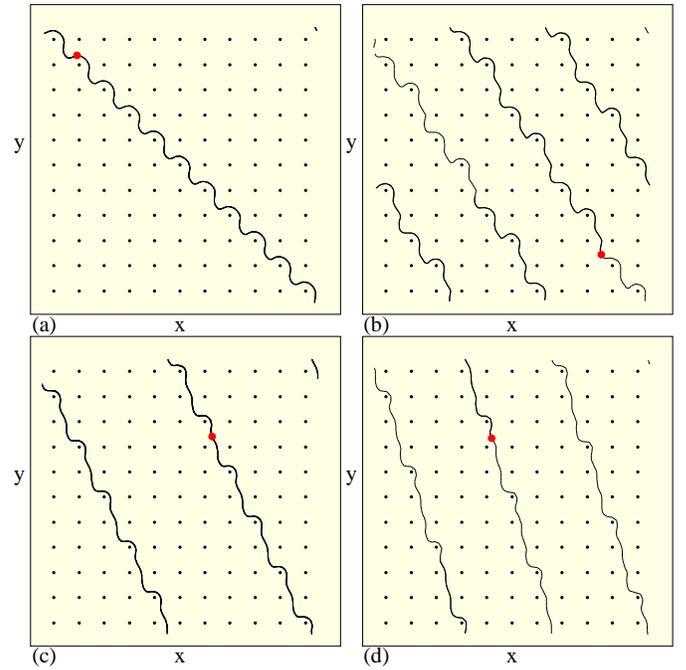}
\caption{
The skyrmion trajectories and substrate maxima.
(a) The $|R|=1/1$ step 
for 
$\alpha_m/\alpha_d = 1.28$ for the system 
in Fig.~\ref{fig:6}(a,c). 
(b) The $|R|=5/3$ step for $\alpha_m/\alpha_d = 1.91$ for
the system in Fig.~\ref{fig:6}(b,d).
(c) The $|R|=2/1$ state for $\alpha_m/\alpha_d = 4.925$ 
for the system in Fig.~\ref{fig:8}(a,c). (d) The $|R|=3/1$ state 
for $\alpha_m/\alpha_d = 4.925$ in Fig.~\ref{fig:8}(a,c).
}
\label{fig:7}
\end{figure}

\begin{figure}
\includegraphics[width=3.5in]{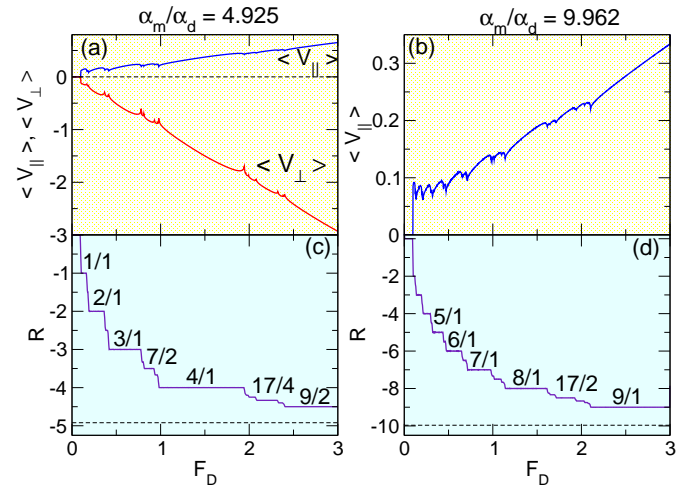}
\caption{
(a) $\langle V_{||}\rangle$ and $\langle V_{\perp}\rangle$ 
vs $F_{D}$ at $\alpha_m/\alpha_d = 4.925$. 
(b) $\langle V_{||}\rangle$ vs $F_D$ for $\alpha_m/\alpha_d=9.962$.
(c) $R$ vs $F_D$ for the sample in panel (a) with 
$\alpha_m/\alpha_d=4.925$.
The steps 
at $|R| = 2/1$, 3/1, 7/2, 4/1, 17/4, and 9/2 are highlighted. 
(d) $R$ vs $F_D$ for the sample in panel (b) with
$\alpha_m/\alpha_d=9.962$, where the steps at
$|R|=5/1$, 6/1, 7/1, 8/1, 17/2, and 9/1 are highlighted.
}
\label{fig:8}
\end{figure}

\subsection{Higher Order Steps}

The specific features in the velocity force curves are strongly dependent on 
the value of $\alpha_m/\alpha_d$.  
In Fig.~\ref{fig:6}(a,b) we show 
$\langle V_{\perp}\rangle$ and $\langle V_{||}\rangle$ 
for $\alpha_m/\alpha_d = 1.28$ and $1.91$, respectively, while
in Fig.~\ref{fig:6}(c,d) we plot the corresponding $R$ vs $F_{D}$ curves. 
For $\alpha_m/\alpha_d = 1.28$, the system is predominantly locked to the  
$|R| = 1/1$ step as shown in Fig.~\ref{fig:6}(b).
On this step, the skyrmion moves along the $45^\circ$ direction
as illustrated in Fig.~\ref{fig:7}(a) for $F_D=1.0$.  Here
the skyrmion follows a sinusoidal trajectory, moving by
one plaquette in the $x$-direction and one
plaquette in the $y$-direction in a single period.  
At values of $F_{D}$ higher than those plotted in Fig.~\ref{fig:6}(a), 
further locking steps occur as $|R|$ approaches $|R| = \alpha_m/\alpha_d=1.28$. 
At $\alpha_m/\alpha_d = 1.91$, a larger number of steps occur,
as indicated in Fig.~\ref{fig:6}(d) where 
we highlight the $|R| = 1/1$, 4/3, 3/2, 5/3, 7/4, 5/9, and $33/18$ steps. 
For $F_D$ values higher than those shown in Fig.~\ref{fig:6}(d),
additional steps 
appear as $|R|$ approaches the clean limit value. 
Figure~\ref{fig:7}(b) illustrates
the skyrmion trajectory on the $|R| = 5/3$ step for the system in 
Fig.~\ref{fig:6}(b).  The skyrmion moves five plaquettes in the direction
perpendicular to the drive for every three plaquettes in the direction
parallel to the drive during a single period.
Fig.~\ref{fig:8}(a) shows $\langle V_{||}\rangle$ and 
$\langle V_{\perp}\rangle$  versus $F_{D}$  
for a sample with $\alpha_m/\alpha_d = 4.925$, and
the corresponding $R$ vs $F_{D}$ curve appears in 
Fig.~\ref{fig:8}(c). 
In this case, the clean limit Hall angle is large, and
there is no longer a phase where the skyrmions move
only in the direction of the  
drive, so the $0/1$ step is lost. 
Instead, above depinning the motion jumps straight into the 
$|R| = 1.0$ state. 
This is followed by steps at 
$|R|=2/1$, 3/1, 7/2, 4/1, 17/4, and $9/2$.
There are also numerous smaller intermediate
steps corresponding to higher order fractions.  
In Fig.~\ref{fig:7}(c) we plot the skyrmion orbit on 
the $|R| = 2/1$ step for the system in 
Fig.~\ref{fig:8}(a,c), while Fig.~\ref{fig:7}(d) shows the skyrmion orbit 
at $|R| = 3/1$ for the same system.  
In general, as $\alpha_m/\alpha_d$ increases, more steps become visible. 
In Fig.~\ref{fig:8}(b) 
we plot only $\langle V_{||}\rangle$ versus $F_{D}$
for a sample with $\alpha_m/\alpha_d = 9.962$ 
to show more clearly the increased number of features in the 
velocity-force curve. 
Figure~\ref{fig:8}(d) shows the corresponding $R$ vs $F_{D}$ curve,
where we highlight the steps at 
$|R| = 5/1$, 6/1, 7/1, 8/1, 17/2, and $9/1$. 
The largest observable integer step has a value of $n$ 
that is the largest integer which is smaller or equal to
the value of
$\alpha_m/\alpha_d$. 

\begin{figure}
\includegraphics[width=3.5in]{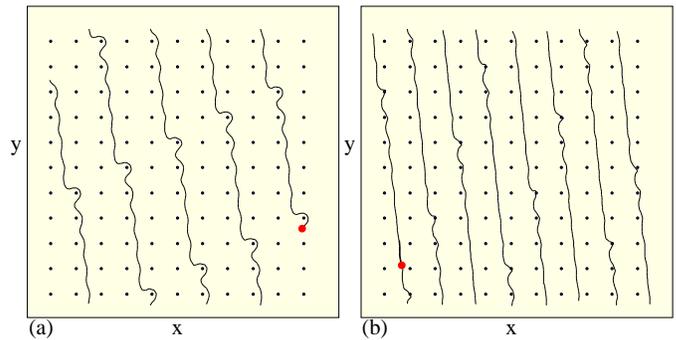}
\caption{Skyrmion trajectories and substrate maxima for
(a) the $|R| = 5.0$ state for the system in Fig.~\ref{fig:8}(b,d) at
$\alpha_m/\alpha_d = 9.962$, and
(b) the $|R| = 8.0$ state for the same system.   
}
\label{fig:9}
\end{figure}

In Fig.~\ref{fig:9}(a) we plot the skyrmion trajectories at 
$|R| = 5.0$ for the system in Fig.~\ref{fig:8}(b,d) with
$\alpha_m/\alpha_d=9.962$, showing that the
skyrmions circle around every fifth substrate maximum.
At $|R|=8.0$, shown in
Fig.~\ref{fig:9}(b),
the trajectories are much straighter 
and the skyrmion moves at an angle that causes it to translate by
eight plaquettes in the negative $y$ direction 
for every one plaquette in the positive $x$-direction. 

\begin{figure}
\includegraphics[width=3.5in]{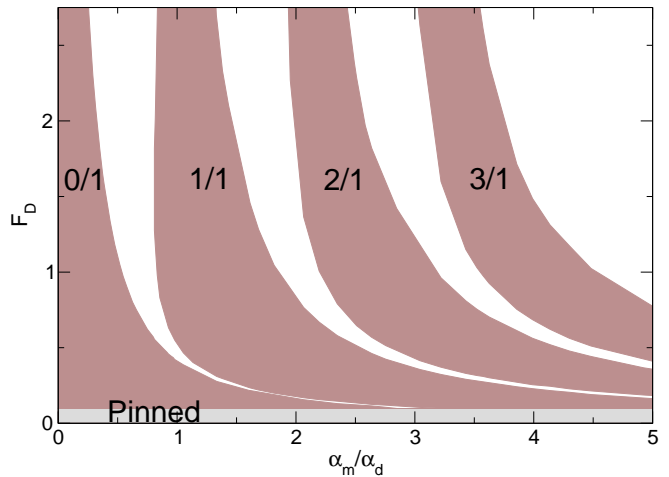}
\caption{
The regions of $F_{D}$ vs $\alpha_m/\alpha_d$ where the pinned phase 
and the $|R|=1/0$, 1/1, 2/1, and $3/1$ locking states occur. 
}
\label{fig:10}
\end{figure}

\begin{figure}
\includegraphics[width=3.5in]{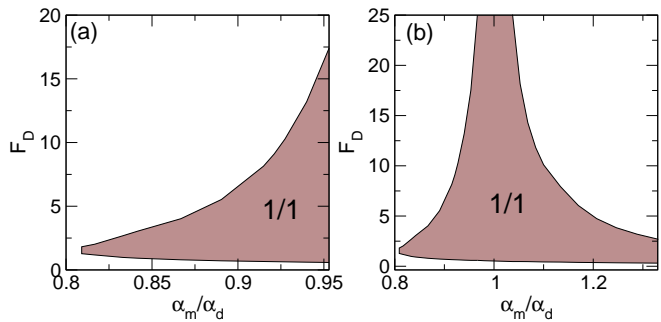}
\caption{(a) $F_{D}$ vs $\alpha_m/\alpha_d$  for 
the $\alpha_m/\alpha_d < 1.0$ side of the $|R|=1/1$ locking step
from Fig.~\ref{fig:10}. (b) 
$F_{D}$ vs $\alpha_m/\alpha_d$ showing only the 
$|R|=1/1$ locking phase from Fig.~\ref{fig:10}.
The width of the step decreases 
for large $F_{D}$.   
}
\label{fig:11}
\end{figure}

\subsection{Arnol'd Tongues}

By conducting a series of simulations for varied $\alpha_m/\alpha_d$ 
we can examine the evolution of the different locking phases. 
In Fig.~\ref{fig:10} we show the evolution of the pinned phase and the 
$|R|=0/1$, 1/1, 2/1, and $3/1$ locking phases 
for $0 \leq F_{D} \leq 3.0$ and $0 \leq \alpha_m/\alpha_d \leq 5.0$. 
The depinning threshold for the 
pinned phase remains roughly constant as a function
of $\alpha_m/\alpha_d$. 
For small $\alpha_m/\alpha_d$, the skyrmion motion 
is strictly in the $x$ direction as indicated by the 
presence of the $|R|=0/1$ phase. 
As $F_{D}$ increases, 
the range and 
the width of the lower order locking phases decreases while new
locking phases appear. 
The regions in which the locking occurs have the characteristic features of 
Arnol'd tongues, which occur in dynamical systems when there are two 
competing frequencies \cite{59,60}.  
In our system, the two frequencies are 
given by the inverse rate of translation along the
x direction and the inverse rate of translation along the y direction
when the system is on a locking step.  These frequencies are quantized
on the step due to the periodicity of the substrate; unlike the case of
Shapiro steps, we do not apply any ac drive and the frequencies arise
from the combination of dc motion and the substrate. If the substrate were
rectangular instead of square, a different set of Arnol'd tongues would arise.
At higher values of $\alpha_m/\alpha_d$, a larger number of locking steps
appear in the velocity-force curves.
For example, at $\alpha_m/\alpha_d= 2.5$, as $F_D$ increases
the system passes through a small region of $|R|=0/1$ 
followed by the $|R|=1/1$ locking phase and then 
by the $|R|=2/1$ phase. 
Another feature of the phase diagram is that at higher drives 
the width of each locking phase decreases. 
To illustrate this more clearly,
we focus on the width of the 
$|R|= 1/1$ step as a function of $F_{D}$, 
as highlighted in Fig.~\ref{fig:11}(a) where we plot
the location of the $|R|=1/1$ step over the narrow range
$0.8 < \alpha_m/\alpha_d < 0.96$ out to $F_D=20$, much higher 
than the maximum value of $F_D$ shown
in Fig.~\ref{fig:10}.
The lower edge of the locking regime, which was dropping
to lower values of $\alpha_m/\alpha_d$ with increasing 
$F_D$ in Fig.~\ref{fig:10},
bends back around for higher $F_D$ as shown in Fig.~\ref{fig:11}, and
shifts to higher $\alpha_m/\alpha_d$ with increasing $F_D$.
The resulting nose structure is the origin of the overshoot phenomenon,
where over a certain range of $\alpha_m/\alpha_d$, the system locks to the
$|R|=1/1$ step for lower $F_D$ only to drop out of that step as $F_D$
increases.  Here the skyrmions can lock to the $|R|=1/1$ direction 
even though their resulting motion follows a higher angle $\Theta$ than
would occur in a clean system.
As $F_{D}$ increases above the edge of the $|R|=1/1$ step,
the skyrmion jumps to smaller values
of $|R|$ in a series of steps.  On each step the skyrmion locks
to different symmetry directions which are closer to 
the clean system value of $|R| = \alpha_m/\alpha_d$. 
The overall shape of the $|R|=1/1$ step is shown in
Fig.~\ref{fig:11}(b), where we plot the range
$0 \leq F_{D} \leq 25$ and $0.8 \leq \alpha_m/\alpha_d \leq 1.32$. 
There is a decrease in the extent of the locking phase at higher 
drives and at higher $\alpha_m/\alpha_d$ values.
The width of the $|R|=1/1$ step gradually decreases and approaches 
a point centered at the $\alpha_m/\alpha_d = 1.0$ value for
the highest drives. 
We find that the regions over which the other integer locking phases 
occur have similar shapes to that shown in Fig.~\ref{fig:11}(b).  

\begin{figure}
\includegraphics[width=3.5in]{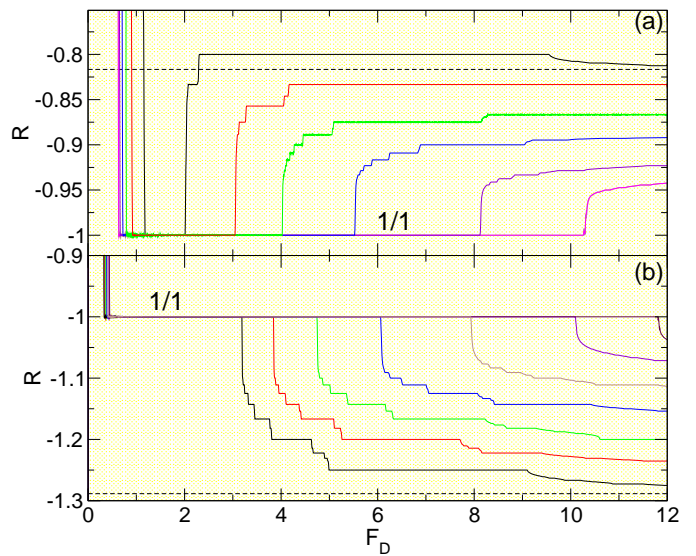}
\caption{(a) $R$ vs $F_{D}$ for 
the $\alpha_m/\alpha_d < 1.0$ regime of the 
$|R|=1/1$ locking phase from Fig.~\ref{fig:11}(b).
From top right to bottom right,
$\alpha_m/\alpha_d = 0.8166$, 0.8418, 0.8668, 0.98041, 0.81475, and $0.9274$. 
The $|R|=1/1$ step is marked,
and the dashed line indicates the value of 
$R$ for $\alpha_m/\alpha_d = 0.8166$ in the clean limit. 
(b) 
$R$ vs $F_{D}$ for the $\alpha_m/\alpha_d > 1.0$ regime of the 
$|R|=1/1$ locking phase from Fig.~\ref{fig:11}(b).
From top right to bottom right, 
$\alpha_m/\alpha_d = 1.084$, 1.1, 1.134, 1.17, 1.207, 1.246, and $1.2885$. 
The $|R|=1/1$ step is marked
and the dashed line indicates the value 
of $R$ for $\alpha_m/\alpha_d = 1.2885$ in the clean limit.  
}
\label{fig:12}
\end{figure}

\subsection{Overshoot Effect for Varied $\alpha_m/\alpha_d$}

A new set of locking steps arises for values of $F_{D}$ 
greater than that where the $|R|=1/1$ locking occurs in Fig.~\ref{fig:11}(a).
In Fig.~\ref{fig:12}(a) we plot $R$ versus $F_{D}$ for different
values of $\alpha_m/\alpha_d$
on the $\alpha_m/\alpha_d < 1.0$ side of the $|R|=1/1$ locking phase 
shown in Fig.~11(b), where an overshoot effect occurs. 
The dashed line is the value 
of $R$ at $\alpha_m/\alpha_d=0.8166$ in
the clean limit; similar overshoots occur for the
other values of $\alpha_m/\alpha_d$. The black curve shows that 
for $\alpha_m/\alpha_d=0.8166$ in the presence of
a substrate,
the skyrmion locks to the 
$|R|=1/1$ direction for $1.2 < F_{D} < 2.0$ and 
then jumps to the $|R| = 4/5 = 0.8$ state. 
For $F_{D} > 9.6$, $R$ jumps off of the $|R| = 0.8$ step and, 
for higher values of $F_{D}$, further small jumps in $R$ occur 
as $R$ approaches the clean value limit. 
As $\alpha_m/\alpha_d$ is increased toward $\alpha_m/\alpha_d = 1.0$, the
width of the interval of $F_{D}$ over which the system is locked 
to the $|R|=1/1$ step increases
since the clean limit Hall angle is closer to the $|R|=1/1$ angle
of $45^\circ$.
In each case,
when the system jumps out of the 
$|R|=1/1$ step for increasing $F_D$, it can jump into a 
series of other locking steps
as $R$ gradually approaches the clean value. 
In Fig.~\ref{fig:12}(b) we 
plot $R$ versus $F_{D}$ for the $\alpha_m/\alpha_d > 1.0$ side of the
$|R|=1/1$ locking phase from Fig.~\ref{fig:11}(b).  
The 
dashed line indicates the clean limit value of $R$ for 
$\alpha_m/\alpha_d = 1.2885$. 
As $\alpha_m/\alpha_d$ approaches $1.0$ from above, the extent of the region
over which the system is locked to the $|R|=1/1$ state again increases.
After jumping out of the $|R|=1/1$ state, the system jumps to
larger values of $|R|$ in a series
of smaller locking steps as it approaches the clean limit 
value of $R$.   
We observe similar changes in $R$ for the other 
integer locking phases such as $|R|=2/1$ and $|R|=3/1$.

\begin{figure}
\includegraphics[width=3.5in]{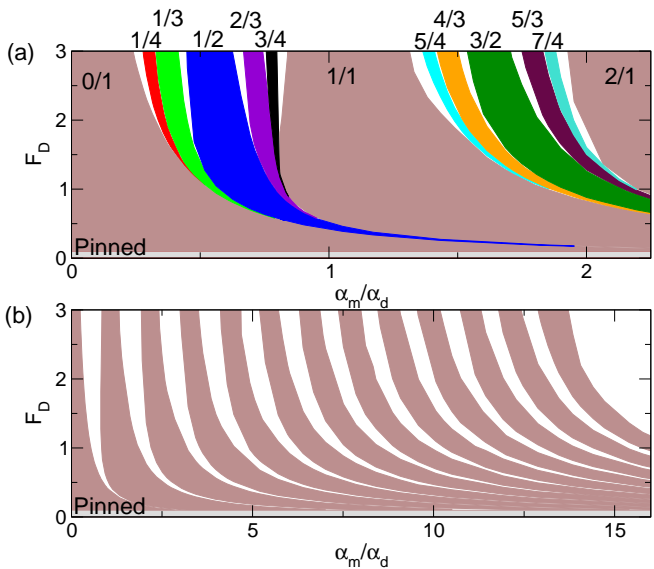}
\caption{
(a) The locking phases for $F_{D}$ vs $\alpha_m/\alpha_d$ 
highlighting the fractional locking steps with
$|R| = 1/4$, 1/3, 1/2, 2/3, 3/4, 5/4, 4/3, 3/2, 5/3, 
and $7/4$ steps along with the integer matching steps
with $|R| = 0/1$, 1/1, and $2/1$. 
(b) Integer locking phases only for $F_D$ vs $\alpha_m/\alpha_d$, for
$|R|=0/1$ through $|R|=12/1$ from left to right.
}
\label{fig:13}
\end{figure}

\subsection{Fractional Locking}

In Fig.~13(a) we plot the locking phases as a function of $F_D$ and
$\alpha_m/\alpha_d$ 
over the range $0 \leq  \alpha_m/\alpha_d \leq 2.5$, 
highlighting the 
fractional 
locking steps at the $|R| = 1/4$, 1/3, 1/2, 2/3, 3/4, 5/4, 4/3, 3/2, 5/3, 
and $7/4$ states falling between the integer steps at 
$|R|=0/1$, 1/1, and 2/1. 
Here the widths of the fractional locking states behave  similarly 
to those of the integer locking states.
Figure \ref{fig:13}(a)
shows that for certain
values of $\alpha_m/\alpha_d$, the fractional steps 
will be the dominant feature observed in transport.
The higher order fractional
steps, not shown in the figure, exhibit similar features. 
The overshoot effect described for the 
$|R|=1/1$ locking step in Fig.~\ref{fig:11} occurs for
all of the integer and fractional locking steps as well, 
with the overall width of each of the steps decreasing 
with increasing $F_{D}$. 
In Fig.~\ref{fig:13}(b) we plot only the integer locking regions 
from $|R|=0/1$ to $|R|=12/1$ as a
function of $F_D$ and $\alpha_m/\alpha_d$
for the range $0 < \alpha_m/\alpha_d  < 16.5$. 
Between each of the integer steps,
there is a series of fractional steps (not shown) similar to that 
illustrated in Fig.~\ref{fig:13}(a). 
The fractional steps have the form $N + n/m$, where $N$ are the integer steps. 

\subsection{Speedup Effects}

\begin{figure}
\includegraphics[width=3.5in]{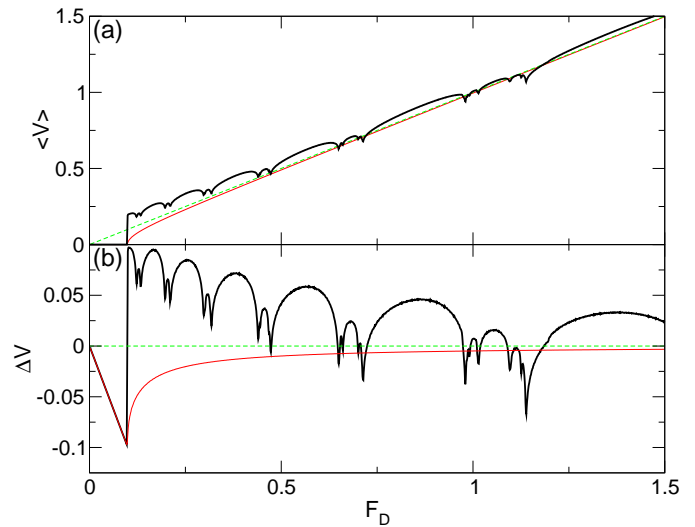}
\caption{(a) The net skyrmion velocity 
$\langle V\rangle = (\langle V_{||}\rangle^2 + 
\langle V_{\perp}\rangle^2)^{1/2}$ 
vs $F_{D}$
for $\alpha_m/\alpha_d = 9.962$ (upper curve), $0.0$ (lower curve), 
and the clean limit value (dashed line).
Here there are regions 
of the $\alpha_m/\alpha_d=9.962$ curve where $\langle V\rangle$ exceeds the
clean limit value, 
indicating 
a speedup or acceleration effect.
(b) $\Delta V = \langle V\rangle - \langle V\rangle_{\rm clean}$ 
for $\alpha_m/\alpha_d= 9.962$ (upper line), $0.0$  
(lower line), and the clean limit value (dashed line). 
The speedup effect is indicated by 
regions in which $\Delta V > 0.$  
}
\label{fig:14}
\end{figure}

The Magnus term can produce an acceleration or speedup effect
of the skyrmion, in which the speed of the skyrmion is higher in the
presence of a substrate than it would be in the absence of a substrate
or in the overdamped limit.
In Fig.~\ref{fig:14}(a) we plot the net skyrmion velocity 
$\langle V\rangle = (\langle V_{||}\rangle^2 + \langle V_{\perp}\rangle^2)^{1/2}$ 
versus $F_{D}$ for 
$\alpha_m/\alpha_d = 9.962$, the overdamped case of $\alpha_m/\alpha_d=0$,
and the clean limit.
In the  overdamped case, 
$\langle V\rangle$ is always smaller than the clean limit value, 
indicating that the substrate does
not accelerate the skyrmion. 
In contrast, for $\alpha_m/\alpha_d = 9.962$ there are clear regions
where $\langle V\rangle$ is {\it higher} than the clean value. 
This effect is most prominent for values of $F_{D}$ 
just above the depinning threshold. 
To show the speedup more clearly,
in Fig.~\ref{fig:14}(b) we plot 
$\Delta V = \langle V\rangle - \langle V\rangle_{\rm clean}$ 
for $\alpha_m/\alpha_d = 9.962$ and 
0.0. 
For the overdamped case, $\Delta V < 0.0$ over the entire range of $F_D$. 
The lowest values of $\Delta V$ occur just at 
depinning, and then $\Delta V$ gradually approaches zero as $F_D$ increases.
The behavior of the overdamped system is similar to the velocity-force
curves observed in other overdamped systems such as 
colloids driven over periodic 
substrate arrays\cite{47,48}.
For $\alpha_m/\alpha_d = 9.962$, $\Delta V$ shows a series of peaks which are
correlated with transitions between the different locking regimes.
For $0.1 < F_{D} < 0.44$,  $\Delta V > 0.0$  due to the speedup effect, where 
there is an enhancement in the net velocity of up to twice the 
velocity in the clean limit. 
As $F_{D}$ is  
further increased the magnitude of the speedup effect 
decreases and there are several intervals of $F_{D}$ where the skyrmion
is moving significantly slower than it would in the clean limit.
For further
increases in $F_{D}$, $\Delta V$  approaches zero. 
The speedup effect is generated by the non-dissipative
terms in the equation of motion
which cause the skyrmions
to be accelerated through certain portions of the substrate potential. 

\begin{figure}
\includegraphics[width=3.5in]{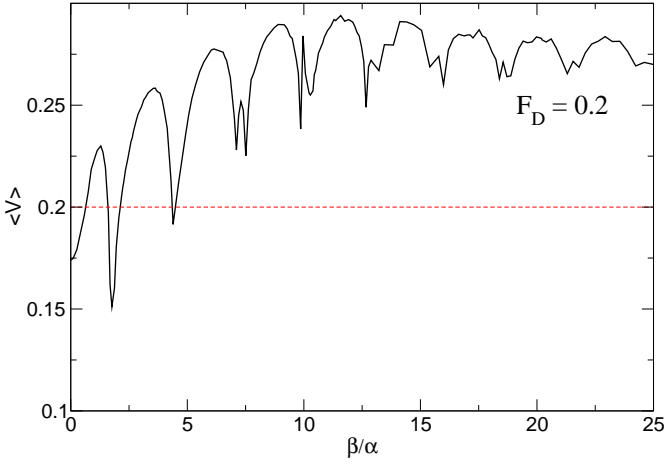}
\caption{ $\langle V\rangle$ vs $\alpha_m/\alpha_d$ for 
fixed $F_{D} = 0.2$. The dashed line indicates
the clean limit value of 
$\langle V\rangle = 0.2$. The speedup effect occurs when 
$\langle V\rangle > 0.2$  
and increases in magnitude with increasing $\alpha_m/\alpha_d$. 
}
\label{fig:15}
\end{figure}

In Fig.~\ref{fig:15} we plot 
$\langle V\rangle$ versus $\alpha_m/\alpha_d$ for a fixed 
$F_{D} = 0.2$, where the dashed line indicates the 
clean limit value of $\langle V\rangle=0.2$. 
Values of $\langle V\rangle > 0.2$ indicate a speedup effect. 
For $\alpha_m/\alpha_d = 0$, in the overdamped limit,
$\langle V\rangle <0.2$. 
We find a
series of oscillations in $\langle V\rangle$
corresponding to the different locking phases through which the
system passes as a function of $\alpha_m/\alpha_d$.
Here, the magnitude of the
speedup effect increases on average with increasing $\alpha_m/\alpha_d$.  

A speedup effect for a driven skyrmion 
interacting with a defect has also been observed in simulations
by M{\" u}ller and Rosch, who find
that for some cases the defect causes a net increase 
in the skyrmion velocity \cite{27}.  
They also find that the magnitude of the speedup 
effect decreases when an externally imposed skyrmion drift velocity
is increased.
This is similar to what we observe for the periodic substrate case 
where for increasing external driving force 
the speedup effect is reduced as shown in Fig.~\ref{fig:14}(b). 
In Ref.~\cite{27}
the speedup was 
up to an order of
magnitude higher than the velocity in the clean limit, while
we observe a velocity enhancement 
of only up to a factor of 2.
This is because in our system there is a minimum critical force
required to depin the skyrmion, whereas in Ref.~\cite{27}
there 
was no lower bound on the imposed drift velocity. 
Since the speedup is increased for lower
drives, M{\" u}ller and Rosch
could access lower drives and obtain larger 
velocity enhancements from the speedup effect. 

\subsection{Two Dimensional Analytic Substrate}

\begin{figure}
\includegraphics[width=3.5in]{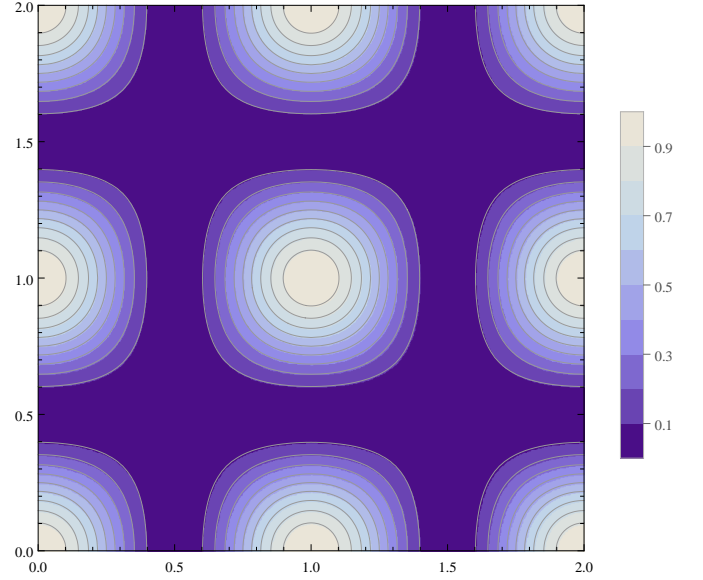}
\caption{ A gray scale image of a portion of the 2D analytic
substrate.  The external drive
is applied along the $x$-direction.   
}
\label{fig:16}
\end{figure}

\begin{figure}
\includegraphics[width=3.5in]{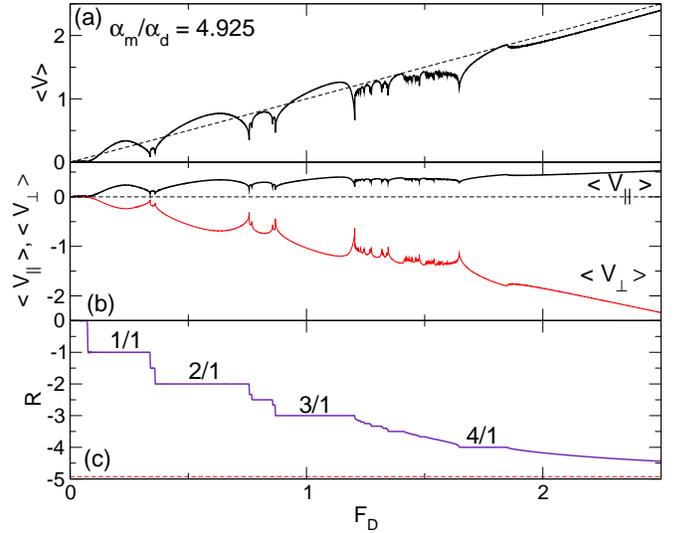}
\caption{The transport for a skyrmion moving over the 
2D sinusoidal substrate illustrated in Fig.~\ref{fig:16} for 
$F_p = 1.5$
and $\alpha_m/\alpha_d = 4.925$. 
(a) The net skyrmion velocity $\langle V\rangle$ vs $F_{D}$ 
shows a series of cusps. 
Dashed line: the clean value limit. 
(b) $\langle V_{\perp}\rangle$ and 
$\langle V_{||}\rangle$ vs $F_{D}$ for the same system.
(c) $R$ vs $F_{D}$ for the same system with the $|R| = 1/1$, 
2/1, 3/1, and $4/1$ steps highlighted. 
There are also numerous smaller scale
fractional locking steps.
}
\label{fig:17}
\end{figure}

In order to understand how general our results are for 
different details of the periodic substrate,
we next consider
a 2D analytic substrate with
the same lattice constant $a$ used in obtaining the previous results. 
The force from the substrate
is given by
$ F_{s}  = F_p[\cos^2(\pi x/a){\hat {\bf x}}
+ \cos^2(\pi y/a){\hat {\bf y}}]$, 
where $F_p$ is the maximum force exerted by the
substrate and $a=3.26$ is the lattice
constant as in the system 
shown in Fig.~\ref{fig:2}. 
Figure~\ref{fig:16} shows 
a gray scale of a portion of the substrate, where the
potential maxima are highlighted.
In Fig.~\ref{fig:17}(a) we plot 
$\langle V\rangle$
versus $F_{D}$ for a skyrmion
moving over a substrate with $F_p = 1.5$
for $\alpha_m/\alpha_d = 4.925$.  The dashed line indicates 
the clean limit value of $\langle V\rangle$.
Figure~\ref{fig:17}(b) shows 
$\langle V_{||}\rangle$ and 
$\langle V_{\perp}\rangle$ vs $F_{D}$ for the same system, 
while in Fig.~\ref{fig:17}(c) we plot the corresponding 
$R$ vs $F_{D}$ curve with a dashed line indicating the
value of $R$ in a clean system and with the
locking steps at $|R| = 1/1$, 2/1, 3/1, and $4/1$ highlighted. 
Although there are some differences
in the details, we find the same general transport features 
for the analytic substrate 
as we observed for the pinned skyrmion substrate
at $\alpha_m/\alpha_d = 4.925$, including the 
locking of the skyrmion motion to different symmetry directions and
the steps in $R$ at integer and fractional ratios of 
$\langle V_{\perp}\rangle/\langle V_{||}\rangle$. 
Figure~\ref{fig:17}(a) 
shows that there are a similar series of dips in 
$\langle V\rangle$ 
corresponding to transitions
between the different locking phases. 
We find that with an analytic potential, 
a somewhat larger number of small looking steps can be resolved,
as shown for $ 1.25 < F_{D} < 1.75$ in Fig.~\ref{fig:17}(a),
and the steps in $R$ in Fig.~\ref{fig:17}(c) for this region 
have a devil's staircase structure. 
Figure~\ref{fig:17}(a) also shows that 
$\langle V\rangle$ exhibits regions where it is 
higher than the clean limit (dashed line), 
indicating
that the same type of speedup effect occurs
for the analytic potential. 
We have also examined the locking effects
on the analytic potential for other values of $\alpha_m/\alpha_d$
and $F_p$, and find that all the features highlighted in 
Fig.~\ref{fig:17} are robust. 
This indicates that the directional locking effect
is a generic feature of skyrmions moving over periodic substrates.           

\section{Scattering Off a Single Pinning Site}

\begin{figure}
\includegraphics[width=3.5in]{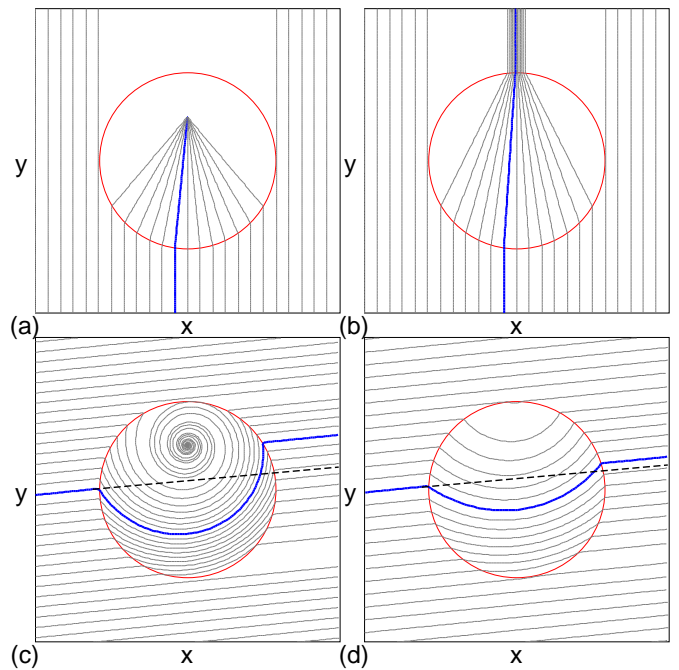}
\caption{ 
Skyrmion trajectories (lines) for a single skyrmion 
at different impact parameters $b$ interacting with a single
pinning site of radius $R_{p} = 0.35$ and maximum pinning force 
of $F_{p}$ for an external drive $F_D$ applied in the 
positive $y$-direction. 
(a) $\alpha_m/\alpha_d = 0.0$, $F_{p} = 0.1$, and  $F_{D} = 0.05$. 
The skyrmions that interact with
the pinning site are captured. 
(b) $\alpha_m/\alpha_d=0$, $F_p=0.1$, and $F_{D} = 0.12$. 
The skyrmions can escape from the pinning site. 
(c) $\alpha_m/\alpha_d = 10.0$, $F_{p} = 0.1$, and $F_{D} = 0.05$. 
The dashed line is a trajectory that a skyrmion would
follow in the absence of the pinning site. 
For a certain range of impact parameters, the skyrmion is captured.
For other impact parameters, skyrmions that escape from the pinning site
have their trajectories shifted as highlighted by the thick line. 
(d) $\alpha_m/\alpha_d=10.0$, $F_p=0.1$, and $F_D=0.12$.
Here the shift in the trajectories of the skyrmions that pass
through the pinning site is reduced and there is less curvature in the 
trajectories.
}
\label{fig:18}
\end{figure}

In order to better understand the dependence of 
the Hall angle on the external drive, we consider the case of
a skyrmion scattering from a single pinning site. 
We drive the skyrmion toward the pinning site
for varied impact parameters $b$, and measure the resulting shift in the
skyrmion position perpendicular to the driving force
for the outgoing state.
We model
the pinning site as a parabolic trap with radius $R_{p}$ 
and a maximum force of $F_{p}$. 
The skyrmion is driven from a point outside the trap towards the trap
with an external drive $F_{D}$ applied in the positive $y$-direction. 
In Fig.~\ref{fig:18}(a) we show the overdamped case with 
$\alpha_m/\alpha_d = 0.0$, $F_{p} = 0.1$, $R_{p} = 0.35$, and $F_{D} = 0.05$, 
We highlight the trajectory of a skyrmion with an 
impact parameter of $b=0.0$.
For this value of $F_{D}$, all skyrmion trajectories that contact the
pinning site
form straight lines directed toward the equilibrium position 
of the pinned state along the $x = 0.0$ line. 
The pinned equilibrium position is
shifted in the positive $y$-direction from the center of the
pinning site due to the applied external force. 
Figure~\ref{fig:18}(b) shows the same system with higher
$F_{D} = 0.12$. 
The trajectories are shifted toward the center of the pinning site,
but since $F_D>F_p$ the skyrmions can escape from the pinning site.
The trajectory for an impact parameter $b=0$ comes in and out of the pinning 
site at $x = 0.0$, so there is no shift, 
while the shift of the trajectories for non-zero impact parameters are 
symmetric across $x=0$.  

Fig.~18(c) shows the trajectories for the same system at
$\alpha_m/\alpha_d = 10.0$ for $F_{D} = 0.05$, where the external drive is
still applied 
in the positive $y$-direction. 
In this case, the trajectories that do not intersect 
with the pinning site move at an angle of
$|\arctan(\alpha_m/\alpha_d)| = 84.3^{\circ}$ with respect to the $y$-axis.
The dashed line represents the trajectory a skyrmion would 
follow in the absence of the pinning site. 
When the skyrmion moves through the pinning site, the trajectory is
no longer straight as in the overdamped case but is now 
strongly curved due to the Magnus force. 
The skyrmions that encounter the pinning site on the upper
left side become trapped by entering a spiral
trajectory that carries them to the
equilibrium pinned position in the upper portion of the pinning site.
Skyrmions that enter the pinning site on the lower left side can escape
from the pin; however,
the position of the outgoing skyrmions are shifted 
in the positive $y$ direction
as indicated by the highlighted trajectory in 
Fig.~\ref{fig:18}(c). 
This shift is similar to the
side jump effect 
that occurs for electron scattering \cite{64}, where the
interaction with pinning or disorder shifts the outgoing trajectory
of the particle relative to its incoming trajectory.
In the case of a periodic substrate,
a skyrmion  
would be repeatedly shifted as it moves through the system, 
so that for $\alpha_m/\alpha_d = 10$ the 
average direction of skyrmion motion
follows a Hall angle that is less than 
the clean value limit of $84.3^{\circ}$. 
Figure~\ref{fig:18}(c) also shows that
when the skyrmion is captured it becomes pinned
along the $x = 0.0$ line at a position above the center of the pin.
The shift of the pinned equilibrium position to this location is caused
by the bias from the external drive, and the location of the equilibrium
position is not changed by the inclusion of a finite Magnus term.
Figure~\ref{fig:18}(d) shows the trajectory for $\alpha_m/\alpha_d = 10$ 
with a higher drive of $F_{D} = 0.12$. 
Here, the size of the side jump 
is reduced, so that the Hall angle that would be observed
for motion through a periodic substrate
is closer to the pin-free value of 
$84.3^{\circ}$. 
This indicates that
the Hall angle increases with increasing driving force, 
as also observed in the simulations with a periodic substrate.
The  shift is reduced at the higher drives in Fig.~\ref{fig:18}(d), 
since the skyrmion is moving more rapidly through the pinning site 
and the trajectory spends less time being bent.

\begin{figure}
\includegraphics[width=3.5in]{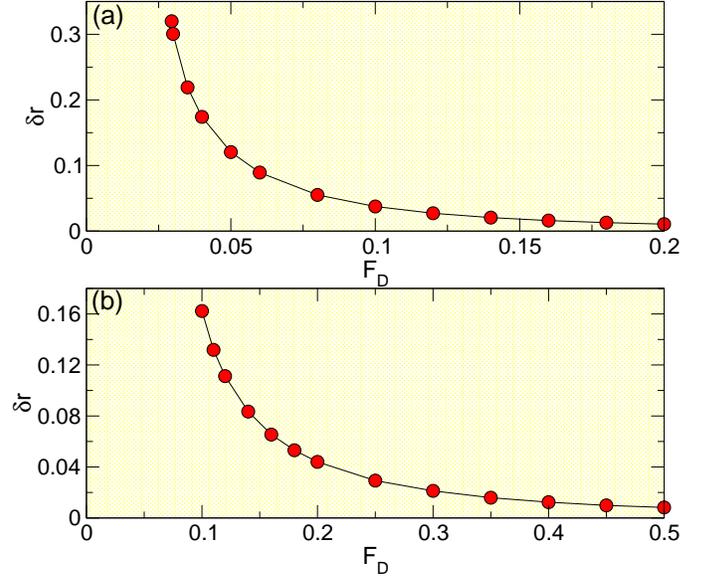}
\caption{
$\delta r$, the magnitude of the trajectory shift, vs $F_{D}$ 
for a skyrmion entering a single pinning site with an impact factor
of zero.
(a) $F_{p} = 0.1$, $R_{p} = 0.35$, and $\alpha_m/\alpha_d = 10.0$. 
$\delta r$ approaches zero at higher drives.
(b) The same for $\alpha_m/\alpha_d = 1.0$.  
}
\label{fig:19}
\end{figure}
    
To quantify the the dependence of the 
magnitude $\delta r$ of the shift or side jump 
of the trajectory as a function of the external drive, in 
Fig.~\ref{fig:19} we plot $\delta r$
for a skyrmion approaching the pinning site with $b=0$.
Figure~\ref{fig:19}(a) shows the shift 
for $F_{p} = 0.1$, $R_{p} = 0.35$, and
$\alpha_m/\alpha_d = 10$. 
The shift is highest at low $F_D$, starting
near $\delta r=0.3$ in Fig.~\ref{fig:19}(a) and gradually approaching
zero as $F_D$ increases.
In Fig.~\ref{fig:19}(b) we plot $\delta r$ versus $F_D$ 
for a sample with $F_{p} = 0.1$, $R_{p} = 0.35$, 
and $\alpha_m/\alpha_d = 1.0$. Here the 
initial value of $\delta r$ is smaller 
due to the smaller value of $\alpha_m/\alpha_d$. 
From an initial value of $\delta r=0.15$, the magnitude of the shift
decreases with increasing drive. 
For $\alpha_m/\alpha_d = 0.0$ (not shown), $\delta r=0$ for all 
values of $F_D$.
If we apply a fit to the decrease of the shift as a function of drive, 
we find 
$\delta r \propto F_{D}^{-\nu}$, 
with $\nu = 1.44-2.0$ depending on the choice of the 
low drive cutoff. 
In the work of M{\" u}ller and Rosch for a skyrmion scattering 
off a single defect \cite{27},  
they found through both simulations and perturbation that the shift as
a function of drive goes as 
$\delta r \propto F_{D}^{-2}$.
Even though the pinning potential
we use has a different form than that used in Ref.~\cite{27},
our results are consistent with the pinning site inducing a trajectory 
shift.      

\begin{figure}
\includegraphics[width=3.5in]{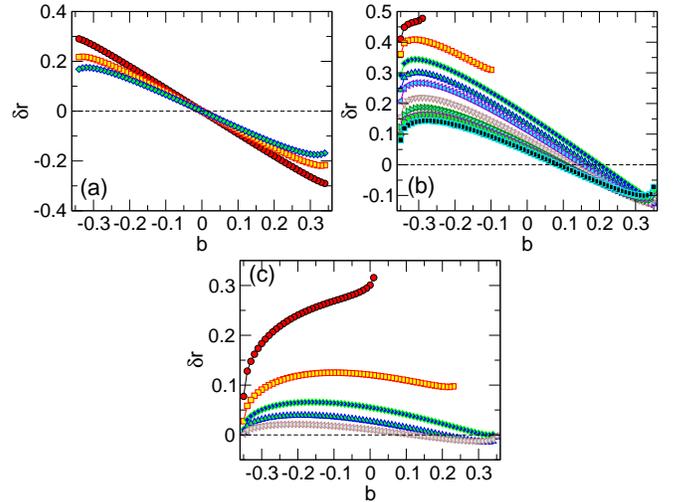}
\caption{
The trajectory shift $\delta r$ vs impact parameter $b$ 
for a skyrmion moving through a pinning site with $F_{p} = 0.1$ 
and $R_{p} =0.35$.
(a) $\alpha_m/\alpha_d = 0$ for $F_D=0.12$, 0.16, and 0.20, from upper left to
lower left.   $\delta r$ is symmetric across $b=0$. 
(b) $\alpha_m/\alpha_d  = 1.0$ 
for $F_D=0.085$, 0.09, 0.10, 0.11, 0.12, 0.14, 0.16,
0.18, and 0.20, from upper left to lower left.
Here the shift is asymmetric across $b=0$. 
For values of $b$ where there are no points, the skyrmion
is captured by the pinning site. 
(b) $\alpha_m/\alpha_d = 10$ 
for $F_D=0.03$, 0.05, 0.08, 0.12, and 0.20, from upper
left to lower left.
The shifts become larger with increasing $\alpha_m/\alpha_d$.
}
\label{fig:20}
\end{figure}

In order to understand the shift of the skyrmion and 
when the skyrmion will be captured by a pinning site,  
in Fig.~\ref{fig:20} we plot $\delta r$
as a function of the impact parameter $b$. 
Figure \ref{fig:20}(a) shows the overdamped case $\alpha_m/\alpha_d = 0.0$ 
for $F_{p} = 0.1$ and $r_{p} = 0.35$ at
$F_{D} = 0.12$, 0.16, and $0.2$.  
At $b=0$, $\delta r=0$ and there is no shift in the trajectory, 
while shifts  
for positive and negative values of $b$ are symmetric across $b=0$
so that the integrated shift over all impact parameters is zero. 
For $F_{D} < F_{p} = 0.1$ the skyrmion is always captured by the pin.  
Figure \ref{fig:20}(b) shows $\delta r$ versus $b$
for a sample with $\alpha_m/\alpha_d = 1.0$ at drives ranging from
$F_{D} = 0.085$ to $F_D=0.20$.  
Here, $\delta r$
decreases with increasing $F_{D}$, 
and the shifts are asymmetric for positive
and negative values of $b$, so that the integrated shift over
all impact parameters is
positive. 
At $F_{D} =  0.085$ and $0.09$, values of $b$ at which there are no points
on the curve indicate that the skyrmion was captured by the pinning site.
For higher values of $F_{D}$, $|\delta r|$ is larger for
$b<0$ than for $b>0$.
The results in Fig.~\ref{fig:20}(b) show that 
increasing $\alpha_m/\alpha_b$ 
reduces the range of impact parameters where an incoming skyrmion is
captured by the pin,
and that for some ranges of $b$ 
a skyrmion can escape the pin even when  $F_{D} < F_{p}$, 
in contrast to the overdamped case where the skyrmion is always trapped
whenever $F_D<F_p$.  

Figure~\ref{fig:20}(c) shows $\delta r$ versus $b$
for $\alpha_m/\alpha_d = 10.0$
over the range $F_{D} = 0.03$ to $F_D=0.2$. 
Here, the ability of the
pinning site to capture the skyrmion is further reduced, and
the skyrmion does not become pinned for more than half the impact 
parameters at $F_{D} = 0.03$ even though 
this drive is substantially smaller than the maximum pinning force $F_p=0.1$.
At $F_{D} = 0.05$ a skyrmion can only
become captured if $b>0.25$. 
The overall shifts are positive over 
a much wider range of $F_{D}$, 
and the magnitude of $\delta r$ decreases with increasing $F_{D}$. 
This result shows that inclusion of a Magnus term in the 
dynamics reduces the ability of pinning sites
to capture particles since the Magnus term induces a side jump or shift
that permits the particles to escape from the pin.

\section{Conclusion} 

We have numerically examined a skyrmion under a dc drive moving over 
a two-dimensional square pinning substrate for
varied ratios of the Magnus force term $\alpha_m$ to the damping $\alpha_d$. 
In the overdamped limit $\alpha_m/\alpha_d = 0$, 
there is a single depinning transition into a sliding state 
where the skyrmion moves in the direction of the applied drive.
For a finite $\alpha_m/\alpha_d$
we find that the skyrmion direction of motion or
Hall angle depends on the magnitude of the external drive and 
gradually approaches
the substrate-free limit at high drives. 
Due to the symmetry of the underlying substrate, the 
Hall angle does not change continuously but 
passes through a series of 
steps as the skyrmion motion becomes locked to certain symmetry 
directions of the substrate. 
These steps occur at integer and rational fractional ratios $n/m$ of the 
perpendicular to parallel velocity components of the skyrmion motion
with respect to the direction of the applied drive,
where $n$ and $m$ are the number of plaquettes the skyrmion moves 
in the perpendicular and parallel directions, respectively.
We find that the Hall angle
generally 
increases with increasing external drive, 
but that there can be an overshoot effect in which the Hall angle is
larger than expected for the substrate-free or clean limit.  In this case,
as the drive increases the Hall angle jumps back to a lower value closer to the
clean limit value.
At the transitions between different directional looking steps, 
the skyrmion velocity shows a striking series of cusps or dips 
where the skyrmion slows down for increasing $F_{D}$, 
producing a negative differential mobility 
at the transitions. 
As $\alpha_m/\alpha_d$ increases, the number of transitions between 
different looking steps increases, as
we map out in a series of phase diagrams. 
The directional locking effects exhibited by the skyrmions 
are very distinct
from the
directional locking effects previously observed for 
overdamped particles such as vortices and colloids interacting with 
periodic substrates.
In the overdamped system, the angle of the external drive must be 
changed with respect to the 
symmetry direction of the substrate in order to induce locking steps,
whereas for the skyrmion system the driving direction remains fixed; only
the magnitude of the driving force is changed. 
Additionally, the overdamped systems exhibit neither
the negative mobility phenomenon at the transitions between
steps nor the overshoot effect. 

We find that the skyrmion motion can exhibit a speedup or acceleration 
effect where the interaction
of the skyrmion with a pinning site can accelerate 
the skyrmion such that the skyrmion velocity is higher than the value that
would 
be induced by the external drive alone.
This effect is generally enhanced at the lower drives and suppressed near the 
directional locking transitions. 
It contrasts strongly with the behavior of overdamped systems, where 
interactions with pinning or a substrate
always decrease the velocity of an overdamped particle.
We find that all of these features are robust for different forms
of the periodic substrate.                 
     
To better understand how the pinning can induce a Hall angle
dependence on the magnitude of the drive, we consider a skyrmion
scattering from a single pinning site. 
In the overdamped limit
with $\alpha_m/\alpha_d=0$, the skyrmion becomes trapped by the pin
whenever the external drive is less than the magnitude of the maximum
pinning force that can be exerted by the pinning site.
For external drives larger than this value, the skyrmion escapes from
the pin and its trajectory is shifted perpendicular to the direction
of skyrmion motion.  The shift is symmetric for positive and negative
impact parameters, so there is no net shift of the trajectory when all
impact parameters are integrated.
In contrast, at a finite $\alpha_m/\alpha_d$, 
the magnitude of the trajectory shift
depends asymmetrically on the impact parameter, producing a 
nonzero net shift of the skyrmion as it moves through
the pinning site when all impact parameters are
integrated.
This shift is similar to the
side jump phenomenon found in electron scattering.  
As the external drive is increased, the net shift decreases and the skyrmion
motion approaches the pin-free trajectory track.
The trajectory shift is responsible for
the modified Hall angle experienced by the skyrmion at low drives,
and the gradual decrease in shift with increasing drive causes the 
Hall angle to gradually approach the clean limit value.
For the case of a periodic substrate, the additional symmetry of the substrate
prevents the angle of the skyrmion motion from changing continuously; instead,
the motion changes in
a series of steps with increasing drive as it approaches the clean limit
value. 
We also find that for finite $\alpha_m/\alpha_d$, depending on the
impact parameter the skyrmion 
may or may not become pinned regardless of whether the maximum pinning force is
smaller than the driving force,
and that for increasing $\alpha_m/\alpha_d$, the range of impact parameters
over which 
the particle is not pinned even when the driving force is smaller than 
the maximum pinning force increases.
This is one of the reasons that pinning of skyrmions is much weaker
than pinning in overdamped systems such as superconducting vortices.

\acknowledgments
This work was carried out under the auspices of the 
NNSA of the 
U.S. DoE
at 
LANL
under Contract No.
DE-AC52-06NA25396.

\end{document}